\documentclass{PoS}

\newcommand{\bea}{\begin{eqnarray}}
\newcommand{\eea}{\end{eqnarray}}
\newcommand{\nn}{\nonumber}
\newcommand{\ra}{\rightarrow}

\title{New applications of CARLOMAT}

\ShortTitle{New applications of CARLOMAT}

\author{\speaker{Karol Ko\l odziej}%
         \thanks{This project was supported in part with financial resources 
of the Polish National Science Centre (NCN) under grant decision 
number DEC-2011/03/B/ST6/01615.}\\
        University of Silesia\\
        E-mail: \email{karol.kolodziej@us.edu.pl}}

\abstract{Modifications of {\tt CARLOMAT}, a program for automatic 
computation of the lowest order cross sections of multiparticle reactions,
that include an abridgement of the phase space integration routine,
an interface to parton density functions, improvement of the color matrix 
computation, supplementation of the Cabibbo-Kobayashi-Maskawa mixing in 
the quark sector, implementation of effective models such as
scalar electrodynamics, the $Wtb$ interaction with operators 
of dimension up to 5 and a general top--Higgs coupling, are discussed.
The modifications, together with recent developments concerning mainly
description of the $e^+e^-$ annihilation into hadrons at low energies,
broaden the spectrum of possible applications of the program.}

\FullConference{Loops and Legs in Quantum Field Theory -- LL2014,\\
		 27 April 2014 - 02 May 2014\\
		 Weimar, Germany}

\begin{document}

\section{Introduction}

{\tt CARLOMAT} \cite{carlomat} is a program for automatic 
computation of the lowest order cross sections of multiparticle reactions,
dedicated mainly to the description of the processes of production and decay
of heavy particles, e.g. top quarks, Higgs boson, or electroweak 
gauge bosons.

Substantial modifications with respect to version 1 of the program that will
be briefly discussed in the subsequent sections of the present lecture 
include: 
\begin{itemize}
\item generation of a single phase space parameterization 
for the Feynman diagrams of the same topology,
\item an interface to parton density functions (PDFs), 
\item improvement of the color matrix computation,
\item the Cabibbo-Kobayashi-Maskawa (CKM) mixing in the quark sector, 
\item effective models, including
scalar electrodynamics, the $Wtb$ interaction with operators 
of dimension up to 5 and a general top--Higgs coupling.
\end{itemize}
Version 2 of {\tt CARLOMAT} was released 
in summer 2013 and the writeup was published in the beginning of 
2014 \cite{carlomat2}.

\section{Phase space integration}

The number of peaks in the squared matrix element usually by far exceeds 
the number of independent variables in a single parameterization of the phase 
space integration element. Therefore, the phase space integration in 
{\tt CARLOMAT} is 
performed with the use of the multichannel Monte Carlo (MC) approach.
In version 1 of the program, a
separate phase space parameterization is generated for each Feynman diagram
and the peaks of the corresponding amplitude, which arise if any Feynman 
propagator approaches its minimum, are smoothed with appropriate mappings of 
the integration variables. The parameterizations are then automatically 
combined in a single multichannel phase space integration routine. 

However, the Feynman diagrams of the same topology may differ from each other 
only in propagators of the internal particles. This means that the integration 
limits 
of all the invariants, that are uniquely defined for each topology in terms
of subsets of four momenta of the final state particles, are the same.
Also the Lorentz boosts of four momenta, which are randomly 
generated in the relative centre of mass system of a given subset of particles, 
to the centre of mass system are the same. 
Thus, both the integration limits
and boosts can be written only once for all the diagrams of the same topology. 
The phase space integration routine in {\tt CARLOMAT\_V2.0}
becomes shorter and the 
compilation time is reduced by a factor of 4--5 for multiparticle reactions,
compared to the previous version of the program.

\section{Hadron--hadron collisions, color matrix and
CKM mixing}

Interfaces to {\tt MSTW} \cite{MSTW} and {\tt CTEQ6} \cite{CTEQ} 
PDFs are added in the MC computation part of the program.
The user should choose if she/he 
wants to calculate the cross
section of the hard scattering process at the fixed centre of mass energy, or
to fold it with the PDFs, treating the initial state
particles as partons of either the $p\bar p$ or $pp$ scattering. 

Computation of the color matrix in {\tt CARLOMAT\_V2.0}
is performed as a separate stage,
that is automatically executed just after the code generation and only
the nonzero elements are transferred to the MC program.
A subroutine {\tt colsqkk} that computes the reduced color matrix 
is divided into smaller subroutines of the user controlled size which 
allows to compute much larger color matrices and speeds up the compilation 
process. 

The CKM mixing in the quark sector is implemented 
in the program. 
However, the complex phase of the CKM matrix can be easily incorporated, as 
the $W$ boson coupling to fermions that always multiplies $V_{ij}$ is complex
anyway. If the CKM mixing is included then the number of Feynman diagrams of 
hadronic reactions grows substantially. 
For the sake of simplicity, only the magnitudes of the CKM matrix 
elements $V_{ij}$ \cite{PDG} are taken into account.
Therefore, as the inclusion of the CKM 
mixing would be an unnecessary complication for many 
applications, it can be either switched on or off.

\section{Anomalous $Wtb$ and top--Higgs Yukawa couplings}

The effective Lagrangian of the $Wtb$ interaction containing operators 
of dimension four and five that is implemented in the current version of
the program has the following form \cite{kane}:
\bea
\label{wtblagr}
L_{Wtb}&=&\frac{g}{\sqrt{2}}\,V_{tb}\left[W^-_{\mu}\bar{b}\,\gamma^{\mu}
\left(f_1^L P_L + f_1^R P_R\right)t 
 -\frac{1}{m_W}\partial_{\nu}W^-_{\mu}\bar{b}\,\sigma^{\mu\nu}
  \left(f_2^L P_L + f_2^R P_R\right)t\right]\nn\\
&+&\frac{g}{\sqrt{2}}\,V_{tb}^*\left[W^+_{\mu}\bar{t}\,\gamma^{\mu}
\left(\bar{f}_1^L P_L + \bar{f}_1^R P_R\right)b
-\frac{1}{m_W}\partial_{\nu}W^+_{\mu}\bar{t}\,\sigma^{\mu\nu}
  \left(\bar{f}_2^L P_L + \bar{f}_2^R P_R\right)b\right].
\eea
The couplings $f_{i}^{L}$, $f_{i}^{R}$, $\bar{f}_{i}^{L}$, $\bar{f}_{i}^{R}$, 
$i=1,2$, can be complex in general. For a detailed explanation of the notation 
used in Eq. (\ref{wtblagr}) see \cite{kkapb}.
In the Standard Model (SM), one has $f_{1}^{L}=\bar{f}_{1}^{L}=1$, while other 
couplings are equal to zero.
If CP is conserved then the following relationships hold:
\bea
\left.\bar{f}_1^{R}\right.^*=f_1^R, 
\quad \left.\bar{f}_1^{L}\right.^*=f_1^L, \qquad
\left.\bar{f}_2^R\right.^*=f_2^L, \quad \left.\bar{f}_2^L\right.^*=f_2^R.
\eea
In order to avoid on-shell poles, masses in the Feynman propagators of 
unstable particles are substituted by:
\bea
\label{m2}
m_b^2 \ra m_b^2-im_b\Gamma_b, \quad b=Z, W, h,\qquad\quad
m_t\ra \sqrt{m_t^2-im_t\Gamma_t},
\eea
in the $s$-, $t$- and $u$-channel.
The top quark width $\Gamma_t$ in (\ref{m2}) is calculated anew every time
the form factors $f_{i}^{L}$, $f_{i}^{R}$, $\bar{f}_{i}^{L}$ and 
$\bar{f}_{i}^{R}$, $i=1,2$, are changed.
For CP-odd choices of the couplings,  
the widths of the top quark $\Gamma_t$ and the width of the antitop quark
$\Gamma_{\bar t}$ calculated using Lagrangian (\ref{wtblagr}) differ
from each other. Therefore
both widths are calculated and the following rule is applied
in the $s$-channel top quark propagators: 
$m_t\ra \sqrt{m_t^2-im_t\Gamma_t}$ is used if the propagator goes into $W^+b$ 
and $m_t\ra \sqrt{m_t^2-im_t\Gamma_{\bar t}}$
is used if the propagator goes into $W^-\bar{b}$ \cite{kkapb}.
The rule does
not work for the propagators in $t$- or $u$-channels, but the actual value
of the width should not play much of a role there.
The prescription was to some extent justified using unitarity arguments
in \cite{kkapb}, but its field theoretical justification would actually 
require the calculation of higher order corrections 
with the nonrenormalizable Lagrangian (\ref{wtblagr}). 

The most general Lagrangian of $t\bar t h$ interaction including corrections
from dimension-six operators that has been implemented in the program has 
the following form \cite{aguilar}:
\bea
\label{tthlagr}
\mathcal{L}_{t\bar t h}=-g_{t\bar th}\bar{t}\left(f
+i f'\gamma_5\right)t h.
\eea
The couplings $f$ and $f'$ in (\ref{tthlagr}) that describe the scalar and 
pseudoscalar departures,
respectively, from a purely scalar top--Higgs Yukawa coupling $g_{t\bar th}$ 
of SM which is reproduced for $f=1$ and $f'=0$, are assumed to be real. 

\begin{figure}
\includegraphics[width=0.9\textwidth]{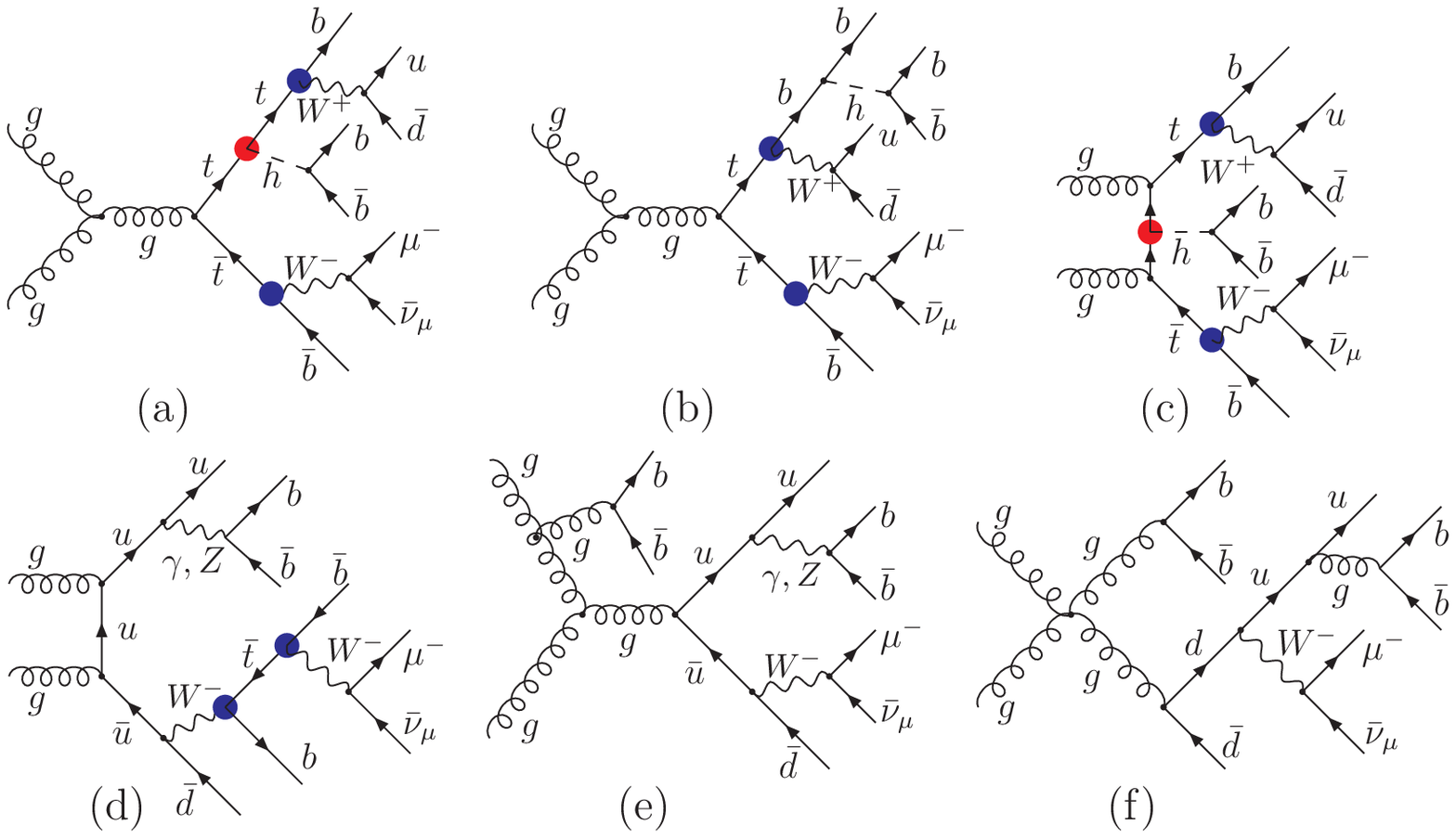}
\caption{Examples of the leading order Feynman diagrams of reaction 
(4.5).}
\label{fig1}
\end{figure}

In order to illustrate the relevance of automation let us consider the
reaction
\bea
\label{ggtth}
gg \;\ra\; b u \bar{d} \bar b \mu^- \bar \nu_{\mu} b \bar b
\eea
which is a dominant partonic subprocess of associated production of the
top quark pair and Higgs boson at the LHC.
Examples of the leading order Feynman diagrams of reaction (\ref{ggtth}) are
shown in Figure \ref{fig1}, where red (blue) blobs indicate the top--Higgs 
$(Wtb)$ coupling.
There are $67\,300$ diagrams in the leading order of the SM,
in the unitary gauge, neglecting masses smaller than the $b$-quark mass 
and the CKM mixing. That big number of the diagrams practically excludes
a possibility of implementing couplings of Eqs. (\ref{wtblagr}) or 
(\ref{tthlagr}) in the matrix element of (\ref{ggtth}) by hand.

The effects of the anomalous top--Higgs Yukawa coupling on different
observables in the process of $t\bar tH$ production at LHC calculated
with {\tt CARLOMAT\_V2.0} were illustrated in \cite{kkjhep}.
The program
was also used to study effects of the anomalous $Wtb$ coupling on
the process of top quark pair production in $p\bar p$ collisions at
the Tevatron \cite{kkplb} and in $pp$ collisions at the LHC \cite{kkapb}.

\section{$e^+e^-$ annihilation to hadrons at low energies}

The knowledge of the energy dependence of the cross section of $e^+e^-$ 
annihilation into hadrons, $\sigma_{e^+e^-\to {\rm hadrons}}(s)$, allows to
determine, through dispersion relations, the hadronic contributions to the 
vacuum polarization which are necessary for better precision of theoretical 
predictions for the muon anomalous magnetic moment and 
evolution of the fine structure constant
from the Thomson limit to high energy scales.

Below  the $J/\psi$ threshold, $\sigma_{e^+e^-\to {\rm hadrons}}(s)$ must
be measured, either by the initial beam energy scan or
with the use of a radiative return method \cite{radret}.
At low energies, the hadronic final states consist mostly of pions, 
accompanied by one or more photons.
The simplest theoretical framework that allows to describe effectively 
the low energetic electromagnetic (EM) interaction of charged pions is  scalar 
electrodynamics (sQED).
At low energies, $\pi^{\pm}$ can be treated as point like particles 
represented by a complex scalar field $\varphi$.
The $U(1)$ gauge invariant
Lagrangian of sQED implemented in {\tt CARLOMAT} has the following form, 
see e.g.
\cite{fred}:
\bea
\mathcal{L}_{\pi}^{\rm sQED}=
\partial_{\mu}\varphi\left(\partial^{\mu}\varphi\right)^*-m_{\pi}^2\varphi\varphi^*
-ie\left(\varphi^*\partial_{\mu}\varphi-\varphi\partial_{\mu}\varphi^*\right)
A^{\mu}+\;e^2g_{\mu\nu}\varphi\varphi^*A^{\mu}A^{\nu}.
\eea
The bound state nature of the charged pion can be taken into account by
the substitutions:
$$ e\to eF_{\pi}(q^2), \qquad e^2\to e^2\left|F_{\pi}(q^2)\right|^2,$$
where $F_{\pi}(q^2)$ is the charged pion form factor that has not been
implemented in the program yet.

Simulation of processes involving the EM interaction of nucleons are
also possible with the most recent version of {\tt CARLOMAT}.
Due to the fact, that the EM current of spin 1/2 nucleons 
has the form
\bea
J^{\mu}&=&e\bar{N}(p')\left[\gamma^{\mu} F_1(Q^2)
+\frac{i}{2m_N}\sigma^{\mu\nu}q_{\nu} F_2(Q^2)\right]N(p),
\eea
which is very similar to the form of the $tb$ current
in $L_{Wtb}$ of Eq. (\ref{wtblagr}), its implementation in the program
was straightforward. Form factors $F_1(Q^2)$ and $F_2(Q^2)$, where 
$Q^2=-(p'-p)^2$, 
have been adopted from PHOKARA \cite{PHOKARA}.

\section{Outlook}

\begin{figure}
\includegraphics[width=0.6\textwidth]{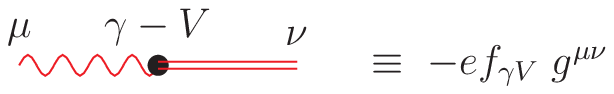}
\caption{A diagram representing mixing of the photon with neutral vector
mesons.}
\label{fig2}
\end{figure}

Work on implementation of the Feynman rules of the Resonance 
Chiral Perturbation Theory provided by Fred Jegerlehner \cite{FJ} is ongoing.
Implementation of new triple and quartic vertices is straightforward.
Just several new subroutines for the calculation of the new
Lorentz tensor structures that arise in the model must be written and
tested. 
However, the implementation of the particle mixing, as illustrated
in Figure \ref{fig2}, is more challenging. It requires substantial changes 
in the code generating part of the program.

\end{document}